\documentclass[12pt,,letterpaper]{JHEP3}
\usepackage{graphics}
\usepackage{epsfig}
\usepackage{slashed}

\title{Holographic superconductors with hyperscaling violation}

\author{ZhongYing Fan\\
Department of Physics, Beijing Normal University, 100875 Beijing,
China\\
\email{zhyingfan@gmail.com}
}

\abstract{We investigate holographic superconductors in asympototically geometries with hyperscaling violation. The mass of the scalar field decouples from the UV dimension of the dual scalar operator and can be chosen as negative as we want, without disturbing the Breitenlohner-Freedman bound. We first numerically find that the scalar condenses below a critical temperature and a gap opens in the real part of the conductivity, indicating the onset of superconductivity. We further analytically explore the effects of the hyperscaling violation on the superconducting transition temperature. We find that the critical temperature increases with the increasing of hyperscaling violation. }

\preprint{}
\begin{document}

\section{Introduction}

In last decades, gauge/gravity duality had been widely used to study condensed-matter theory (AdS/CMT) with great advantage. One of remarkable achievement is the holographic models of superconductor \cite{1,2,3} by employing the Higgs mechanism in the bulk. The black hole becomes unstable below a critical Hawking temperature such that the scalar condensate develops to stabilize the system against small perturbations. Since the gravity is weakly coupled, the dual superconductor is strongly coupled, behaving qualitatively different from the BCS type of superconductors. This is very interesting since the conventional BCS theory describes weakly coupled superconductors well but fails in the strongly coupled one which is however nicely studied in holography.

Generally, the geometry is asymptotically AdS which is dual to relativistically conformal theory in the boundary. Since the condensed-matter theory is non-relativistic in most cases, it is of great importance to develop the corresponding non-relativistical holography. This is achieved in Lifshitz-like geometry with dynamical exponent\cite{4,5,6,7}, where the time scales different from the space, compatible with the peculiar behavior near the quantum critical point in condensed-matter physics. It was further generalized to include hyperscaling violation, another important exponent in low energy physics of condensed-matter theory using the standard Einstein-Maxwell-dilaton gravity model in the bulk\cite{8,9,10,11,12,eoin}. From the perspective of application to realistic systems, it is of certain interests to explore the effects of dynamical exponent and hyperscaling violation on the holographic superconductors. The former had been done in \cite{13} while the latter will be investigated in this paper. Related work also appears in \cite{alberto2012,alberto2013,momeni}.

In geometries with hyperscaling violation, the dual theory is not scale invariant, qualitatively different from the cases in AdS-Lifshitz black hole. Moreover, the hyperscaling violation emerges, generally speaking, just below some non-trivial scale parameter dual to the deep bulk interior. But we will simply assume the unknown parameter is of order of the UV cut-off, extending the geometries with hyperscaling violation to the boundary. Despite this subtlety, we will numerically show that the basic properties of holographic superconductors in this background is as well as those in AdS-Lifshitz black hole. Below a critical temperature the scalar operator condenses but diverges in the zero temperature limit. The real part of the conductivity monotonously approaches 1 in the high frequency limit and becomes exponentially small near zero frequency region, indicating that a gap opens. Moreover, there exists a pole at zero frequency in the imaginary part of the conductivity which implies that a delta peaks emerges in the real part of conductivity from Kramers-Kr\"{o}nig relation.

The remainder part of this paper is organized as follows: In section 2, we briefly review the gravity model of geometries with hyperscaling violation. In section 3, we derive the non-linear equations of motion, solve them numerically and explore the critical temperature of the superconductor using Sturm-Liouville method, extracting the effects of hyperscaling on the critical temperature. Finally, we present a short conclusion in section 4.

\section{Preliminary}

The black hole solution with hyperscaling violation reads\cite{4,11}

\begin{equation}  ds_{d+2}^2=\frac{R^2}{r^2} r^{2\theta/d} (-r^{-2(z-1)}h(r)dt^2+h(r)^{-1}dr^2+dx_i^2) \label{metric} \end{equation}

\begin{equation} h(r)=1-(\frac{r}{r_+})^{d+z-\theta}  \label{metric2}\end{equation}

where $R$ is AdS radius, $d$ is the spatial space dimension, $z$ is dynamical exponent, $\theta$ is the hyperscaling violation exponent, $r_+$ is the location of the event horizon and $h(r)$ is the thermal factor. The Hawking temperature and the entropy density are given by

\begin{equation} T=\frac{1}{4 \pi} \frac{d+z-\theta}{r_+^z}\ ,\qquad S_{en}=\frac{R^d}{4G_{d+2}}\frac{1}{r_+^{d-\theta}}  \label{temp}\end{equation}

where $G_{d+2}$ is the Newton constant in $d+2$ dimensional spacetime. Note that $S_{en}\sim T^{(d-\theta)/z}$, establishing that $\theta$ is the hyperscaling violation exponent. In the asymptotically boundary $r\rightarrow 0$, $h(r)\rightarrow 1$, the metric (\ref{metric}) reduces to

\begin{equation}  ds_{d+2}^2=\frac{R^2}{r^2} r^{2\theta/d} (-r^{-2(z-1)}dt^2+dr^2+dx_i^2) \label{metric3} \end{equation}

which transforms as

\begin{equation}\label{scale}
t \rightarrow \lambda^z t,\ x_i\rightarrow \lambda x_i,\ r\rightarrow \lambda r,\ ds\rightarrow \lambda ^{\theta/d} ds\end{equation}

Evidently, this metric is not scale invariant, quite different from the AdS-Lifshitz invariant background. When $\theta=0$, it reduces to the pure Lifsthize case. The black hole solution (\ref{metric}-\ref{metric2}) is manifestly constructed in the standard Einstein-Maxwell-dilaton (EMD) model

\begin{equation} S=\int \mathrm{d}^{d+2}x \sqrt{-g} ({\Re -2 (\nabla \Phi)^2- Z(\Phi) \mathcal{F_{\mu\nu}}\mathcal{F^{\mu\nu}}}-V(\Phi)) \end{equation}

where $\Phi$ is the dilaton field, $\mathcal{F_{\mu\nu}}$ is an Abelian gauge field. The functions $Z(\Phi)$ and $V(\Phi)$ are generally taken as exponential forms $Z(\Phi)\sim e^{\alpha \Phi}$, $V(\Phi)\sim e^{\beta \Phi}$. We drop the full matter solutions of the EMD theory, since they are irrelevant in our following discussions. One can refer to literatures for details if having interests (as an example, see ref.\cite{8}).

\section{Holographic superconductors with hyperscaling violation}

Before discussing superconductors with hyperscaling violation, we point out that in metric (\ref{metric}), to admit a stable theory, constrained by the null energy conditions or equivalently the real of the dilaton solution, and to extract the dimension of operators as in the standard holographic procedure, the dynamical exponent and hyperscaling violation should satisfy the following conditions

\begin{equation} z\geq 1+\frac{\theta}{d}\ ,\quad 0\leq \theta < d \label{bf}\end{equation}

We will focus on $d=2,\ z=2,\ \theta > 0$ case in the sections below which appears frequently and very interesting in condensed-matter theory.

\subsection{Equations of Motion}

We start from the Abelian-Higgs model in the bulk. The action reads\footnote{Note that the gauge field in the Higgs sector is different from the one coupled to the dilaton which actually diverges in asymptotic limit, leading to no well definition for the finite density. The superconducting instability for the scalar operator charged under the latter gauge field has been investigated with great detail in ref.\cite{alberto2012,alberto2013} where the hyperscaling violation emerges in the deep IR. It is different from our case. The gauge field in eq.(2.6) is simply used to generate asymptotical geometries with hyperscaling violation and remains unbroken throughout this paper.}

\begin{equation} S_{Higg}=\int \mathrm{d}^4 x \sqrt{-g} (-\frac 14 F_{\mu\nu}F^{\mu\nu}- |\partial_\mu \Psi- i q A_\mu \Psi|^2-m^2 |\Psi|^2)  \label{higgs}\end{equation}

And we will work in the probe limit, i.e. the matter fields above don't backreact on the background. For convenience, We set $R=q=1$ and require the matter
solutions are of the following type
 \begin{equation} A=\phi(r)dt\ ,\quad \Psi=\psi(r) \end{equation}

By variation of the action with respect to the scalar and gauge field, we obtain

\begin{equation} \psi''+[\frac{h'(r)}{h(r)}-\frac{\Delta-1}{r}]\psi'+[\frac{\phi^2 r^2}{h^2(r)}-\frac                                                                                                 {m^2 r^{\theta-2}}{h(r)}]\psi=0 \label{eom1}\end{equation}

\begin{equation} \phi''+\frac 1r \phi'-\frac{2\psi^2 r^{\theta-2}}{h(r)}\phi=0\label{eom2}  \end{equation}

where $\psi(r)$ is real, the overall phase factor chosen to zero, allowed by the Maxwell equation. And $\Delta=4-\theta$ is the dimension of the scalar operator. Notice that the mass square of the scalar field decouples from the dual operator dimension. Instead, the hyperscaling violation appears in the UV dimension. Thus, the unitarity bound is always satisfied under the condition (\ref{bf}) such that the mass square can be chosen as negative as one needs. This is the only counter example we know in holography. As is known to all, in the standard AdS/CFT correspondence, the boundary operator is dual to the bulk field and the operator dimension is given by the field mass. However, in our background asymptotically with hyperscaling violation, the situation is qualitatively different. Unfortunately, a better understanding is still lacking\footnote{One possible clue is that the dual boundary theory in our background is not scale invariant from the UV fixed point at the beginning. This is the crucial difference from the theories in the standard AdS/CFT. The dynamical scale below which the hyperscaling violation emerges is the same order of the UV cut-off which may be the origin of the non-trivial behavior of the mass of the bulk fields. It should be emphasized that this result will not change if the hyperscaling violating geometry was embedded into an AdS asymptotical background, in which the mass square of the scalar field cannot be arbitrary negative due to the Breitenlohner-Freedman bound.}.

In order to discuss the conductivity, we need to turn on small perturbations on the Maxwell field. Due to the rotational invariance of the background, we can simply choose the perturbative mode along only one spatial direction

\begin{equation} a_x(t,r,\vec{x})=e^{-i \omega t} a_x(r) \end{equation}

where the spatial momentum has been set to zero for convenience to calculate the AC conductivity. The equation of motion for $a_x(r)$ is given by

\begin{equation} a_x''+[\frac{h'(r)}{h(r)}-\frac 1r]a_x'+[\frac{\omega^2 r^2}{h^2(r)}-\frac{2\psi^2 r^{\theta-2}}{h(r)}]a_x=0  \label{eom3}\end{equation}

Under the coordinate transformation $u=r/r_+$, above equations of motion can be expressed as follows

\begin{equation}\label{eom11}
\psi''+[\frac{h'(u)}{h(u)}-\frac{\Delta-1}{u}]\psi'+[\frac{\phi^2 r_+^4 u^2}{h^2(u)}-\frac{m^2 r_+^{\theta}u^{\theta-2}}{h(u)}]\psi=0  \end{equation}

\begin{equation}\label{eom22} \phi''+\frac 1u \phi'-\frac{2r_+^{\theta}\psi^2 u^{\theta-2}}{h(u)}\phi=0   \end{equation}

\begin{equation}\label{eom33} a_x''+[\frac{h'(u)}{h(u)}-\frac 1u]a_x'+[\frac{\omega^2 r_+^4 u^2}{h^2(u)}-\frac{2r_+^{\theta}\psi^2 u^{\theta-2}}{h(r)}]a_x=0   \end{equation}

where prime now denotes the differentiation with respect to $u$. $u=1, 0$ corresponds to the location of the horizon and the boundary respectively.

\subsection{Boundary conditions}

To obtain physically sensible solutions, we need to impose proper conditions at the horizon and the boundary, which corresponds to specify the in and out states in the dual theory. At the event horizon, the scalar field $\psi(u)$ should be a constant and the temporal component of the gauge field $\phi(u)$  vanishes due to the finite norm of the gauge potential. Furthermore, in order to allow the extraction of the retarded correlators, in-falling conditions required for the perturbative mode $a_x(u)$. Hence we obtain

\begin{equation} \phi(1)=0\ ,\quad \psi(1)=const\ ,\ a_x(u)|_{u\rightarrow 1}\sim (1-u)^{-i\omega/4 \pi T} \label{bdy1}\end{equation}

At the asymptotically boundary, the leading behaviors of the fields are given by

\begin{equation} \psi(u)|_{u\rightarrow 0}=\psi_0+\psi_1 r_+^{\Delta} u^{\Delta}  \label{bdy2}\end{equation}

\begin{equation} \phi(u)|_{u\rightarrow 0}=-\rho+\mu \log{u}  \label{bdy3} \end{equation}

\begin{equation} a_x(u)|_{u\rightarrow 0}=a_x^{(0)}+a_x^{(1)}r_+^2 u^2 \label{bdy4} \end{equation}

where several notes need to be explained. First, since the mass square of the scalar field decouples from the UV dimension of the dual operator, which is peculiar in asymptotically geometries with hyperscaling violation\cite{14}, there is only one operator dual to $\psi_1$ having non-trivial dimension for any fixed mass. The other one with zero dimension is a source in the boundary which should be vanish to give a stable theory

\begin{equation} \psi_0=0 \quad \mathrm{and} \quad \langle O_{\Delta}\rangle= \psi_1 \label{res1}\end{equation}

Notice that the Breitenlohner-Freedman bound has been automatically satisfied under the condition (\ref{bf}) and the mass square can be chosen as negative as we want which is quite different from the case in AdS-Lifshitz background. In eq.(\ref{bdy3}), $\rho$ and $\mu$ are the charge density and the chemical potential of the dual boundary theory, respectively. The logarithmic behaivor of the function $\psi(u)$ originates from the special dynamical exponent $z=2$ and disappears when $z\neq 2$. Finally, by using Kubo formula, the conductivity can be expressed as

\begin{equation} \sigma(\omega)=\frac{1}{i\omega} \frac{a_x^{(1)}}{a_x^{(0)}}   \label{res2}\end{equation}

\subsection{Numerical results}

Having presented so much preparation in the subsections above, we are readily to solve the equations of motion (\ref{eom11})-(\ref{eom33}) with boundary conditions (\ref{bdy1})-(\ref{bdy4}) to show the onset of superconductivity of the system below some critical temperature. Since the equations of motion are non-linearly coupled, we will first numerically solve them to show the full phase diagram for the scalar condensate and the conductivity over the whole temperature and frequency space.

\FIGURE[ht]{\label{figure1}
\includegraphics[width=7.5cm]{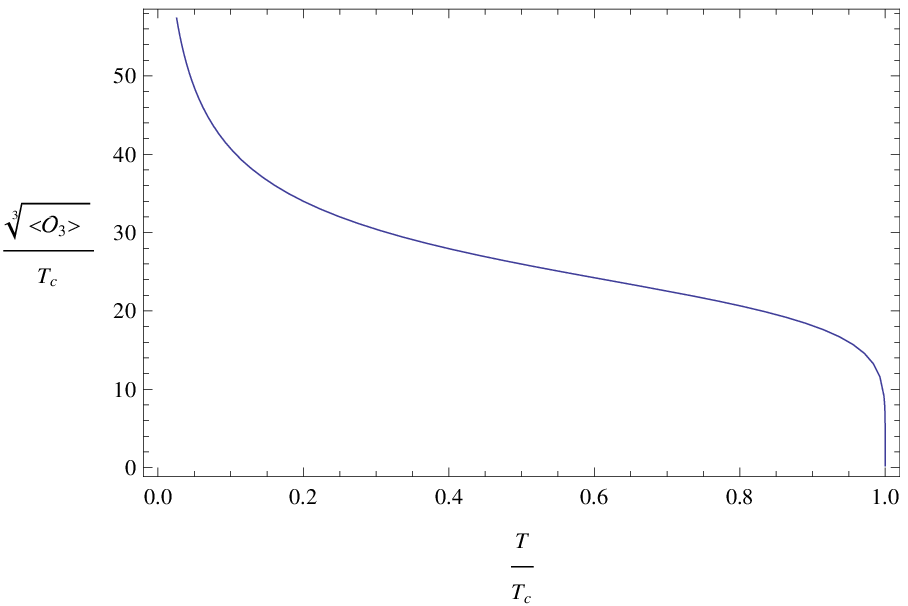}
\includegraphics[width=7.5cm]{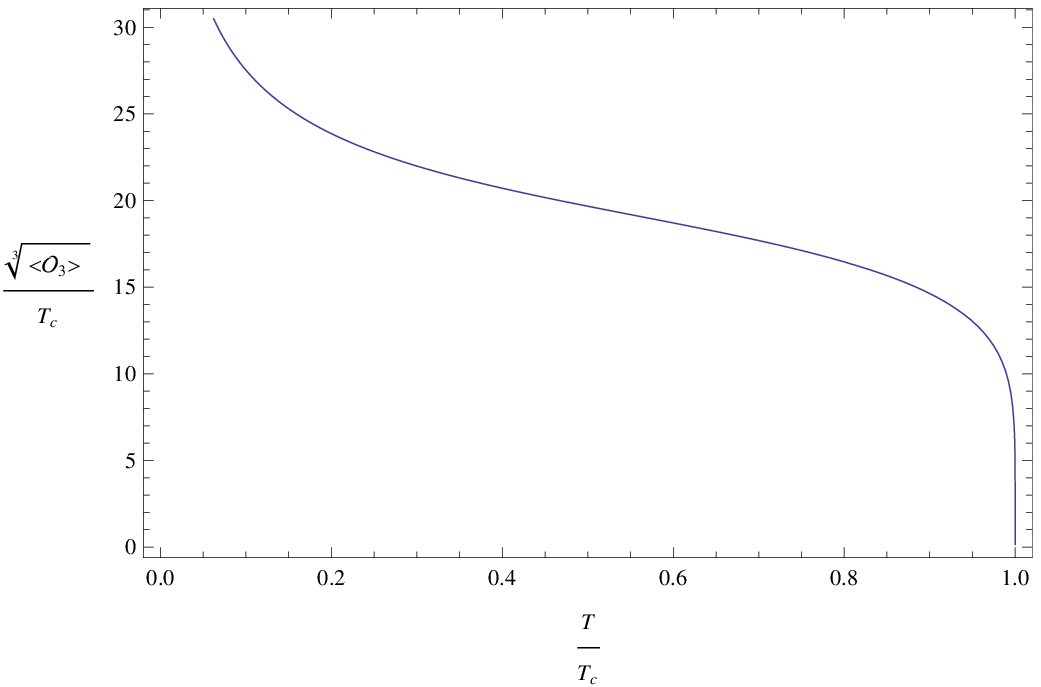}
\caption{The plots of the scalar condensate $\sqrt[3]{\langle O_3 \rangle}/T_c$. The left plot for $m^2=0$, the right plot for $\tilde{m}^2=-3$.}
}
For convenience, we set $\theta=1$. Without loss of generality, we choose two different value for mass square $m^2=0$ and $\tilde{m}^2=-3$, where $\tilde{m}^2=m^2 r_{+c}^{\theta}$ %\footnote{This combination is taken to ensure the numerical calculation is manageable. In the numerical code, we use a more convenient coordinate $u$ defined by $u=r/r+$ and vary the scalar hair to denote the varying of the temperature. This is well known in the codes setting of holographic superconductors.}%
, $r_{+c}$ is the critical radius of the horizon dual to the critical temperature.

From figure \ref{figure1}, we can see that the scalar condensate happens when the temperature below a critical value $T_c$ but doesn't approach a fixed constant in the low temperature limit. In fact, it diverges as $T^{-0.55}$ for $m^2=0$ and $T^{-0.58}$ for $\tilde{m}^2=-3$ by fitting our numerics respectively. This is same as the weakly coupled BCS type superconductors and strongly coupled holographic superconductors in AdS black hole background. In the zero temperature limit $T\rightarrow 0$, the system clearly breaks down and the backreaction effects cannot be neglected any longer.

Near the critical temperature, we also find that the mean-field behavior holds in our numerical result

\begin{equation} \langle O_3 \rangle=(27.671 T_c)^3 (1-T/T_c)^{1/2}\ ,\quad \mathrm{for}\ m^2=0 \end{equation}

with the critical temperature $T_c=0.0229604 \mu$. And

\begin{equation} \langle O_3 \rangle=(21.6728 T_c)^3 (1-T/T_c)^{1/2}\ ,\quad \mathrm{for}\ \tilde{m}^2=-3 \end{equation}

with the critical temperature $T_c=0.0334913 \mu$.

\FIGURE[ht]{\label{figure2}
\includegraphics[width=7.5cm]{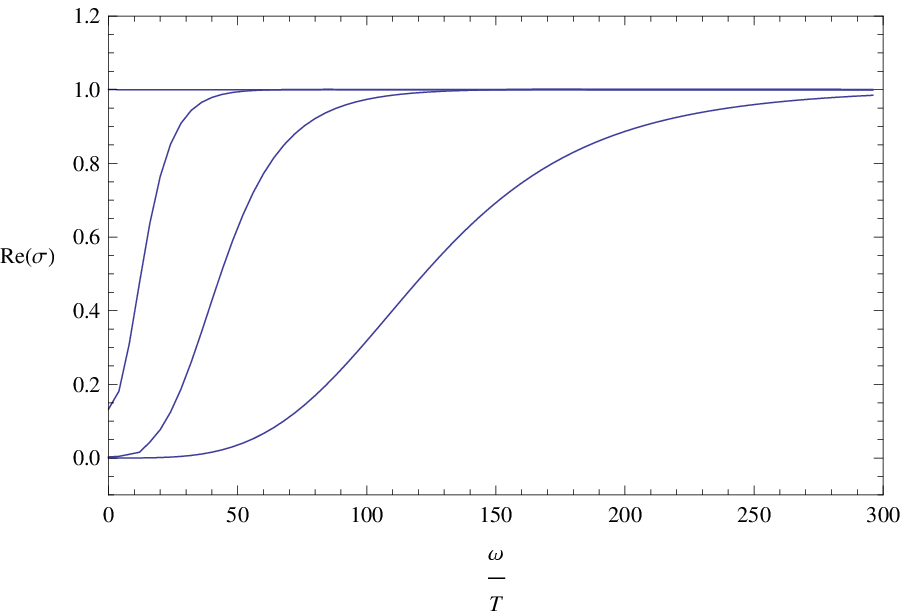}
\includegraphics[width=7.5cm]{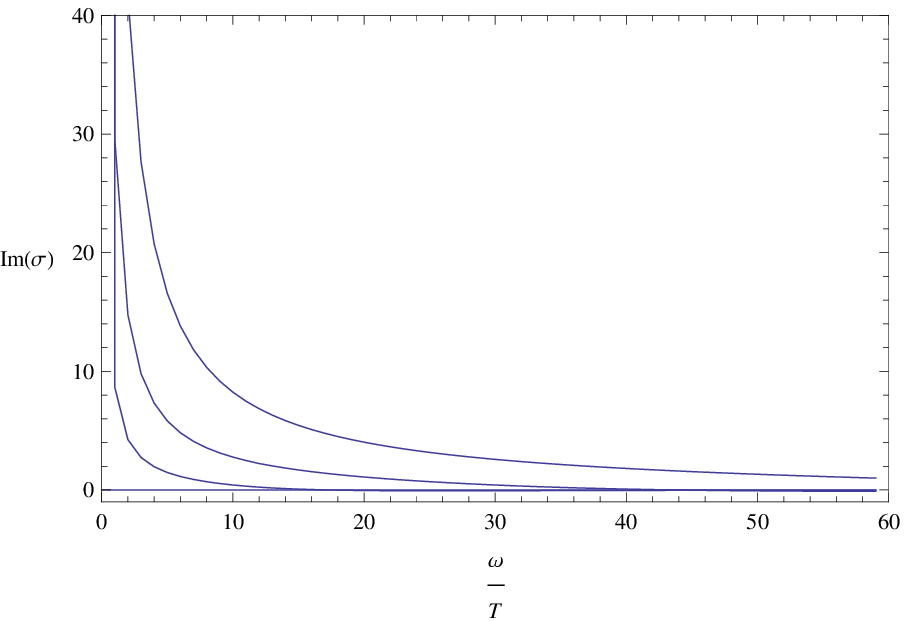}
\includegraphics[width=7.5cm]{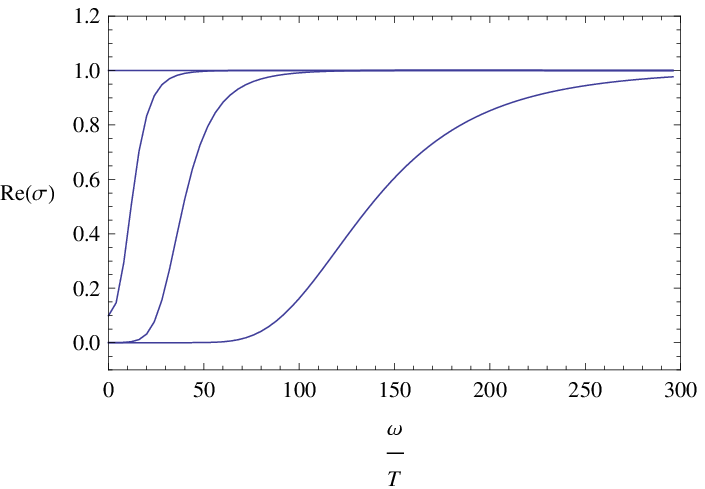}
\includegraphics[width=7.5cm]{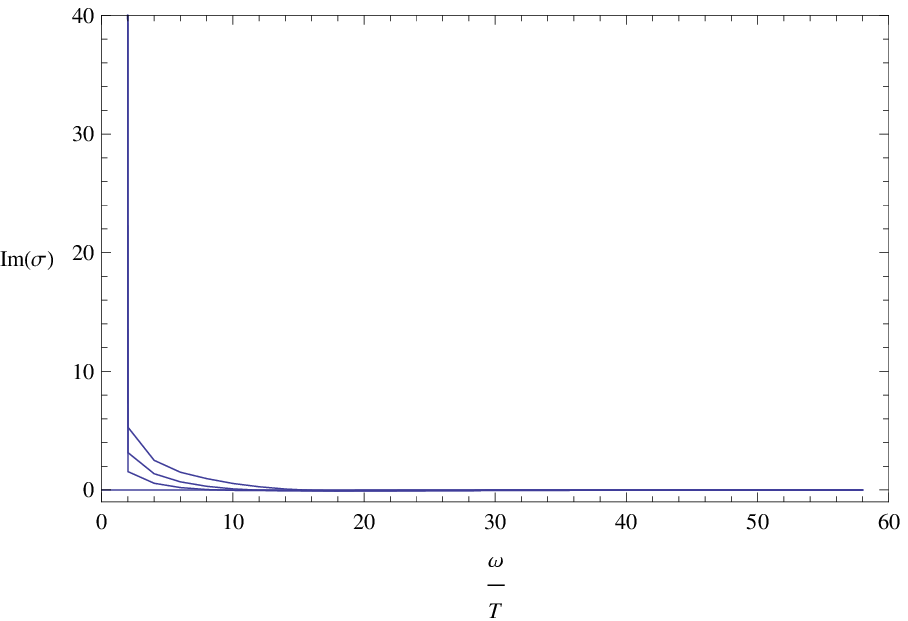}
\caption{The real and imaginary part of the conductivity. The plots above for $m^2=0$ and $T/T_c=1,\ 0.66355,\ 0.275923,\ 0.103421$ from top to bottom. The plots below for $\tilde{m}^2=-3$ and $T/T_c=1,\ 0.673797,\ 0.261434,\ 0.070341$ from top to bottom, respectively.}
}

In figure \ref{figure2}, we plot the real and imaginary part of the AC conductivity for $m^2=0$ and $\tilde{m}^2=-3$. From this figure, we find that the real part of the conductivity approaches 1 in the high frequency limit, consistent with the normal phase and becomes exponentially small near the zero frequency region, indicating that a gap opens. The gap increases deeper as the temperature is lowed and actually diverges in the zero temperature limit. Furthermore, there exists a pole in the imaginary part of the conductivity, $Im[\sigma(\omega)]\sim 1/\omega$, implying that a delta function is contained in the real part of the conductivity $Re[\sigma(\omega)]\sim \pi \delta(\omega)$ according to the Kramers-Kr\"{o}nig relation

\begin{equation} Im[\sigma(\omega)]=-P\  \int_{-\infty}^\infty \frac{\mathrm{d}\omega'}{\pi} \frac{Re[\sigma(\omega)]}{\omega'-\omega} \end{equation}

The delta peak at the zero frequency of the real part of AC conductivity is exactly the infinite DC conductivity, which is compatible with the intuition of superconductivity\footnote{In a translational invariant background, there exists also a delta peak at zero frequency in the real part of the conductivity at normal state. However, the peak will disappear in the probe limit where the translational invariance is effectively broken. For details, see \cite{2}.}.

\subsection{Analytic approach}

To further explore the effects of the hyperscaling violation on the holographic superconductors, we need to vary $\theta$ in the full parameter space eq.(\ref{bf}) to search the possibly underlying imprint on the superconducting phase transition. However, when $\theta$ is fractional, it is very difficult to solve the equations of motion in numerical ways. Hence, analytic approach is indispensable and will show its power in this little subject.\\
The main property we focus on is the dependence of the critical temperature $T_c$ on the hyperscaling violation $\theta$. Thus, we only need to solve the equations of motion (\ref{eom11})-(\ref{eom33}) with boundary conditions (\ref{bdy1})-(\ref{bdy4}) analytically near the phase transition where the scalar condensate remains still very small and can be treated as an expansion parameter. For convenience, we set

\begin{equation} \epsilon \equiv \langle O_\Delta \rangle  \end{equation}

Near the critical point, the scalar field $\psi$ and the gauge field $\phi$ can be expanded as follows\cite{15,16,17}

\begin{equation} \psi=\epsilon \psi_1+\epsilon^3 \psi_3+...  \nonumber\end{equation}

\begin{equation} \phi=\phi_0+\epsilon^2 \phi_2+... \label{expansion}\end{equation}

where $\epsilon \ll 1$. From this expansion, the chemical potential is allowed to be corrected order by order

\begin{equation} \mu=\mu_0+\epsilon^2 \delta \mu_2+... \end{equation}

where $\delta \mu_2>0$. Thus, near the phase transition the scalar condensate can be expressed as

\begin{equation} \epsilon=(\frac{\mu-\mu_0}{\delta \mu_2})^{1/2} \end{equation}

The exponent $1/2$ is the universal result of the mean-field theory, consistent with our numerics in the previous section. Evidently, when $\mu \rightarrow \mu_0$, the order parameter vanishes and the phase transition can happen, implying that $\mu_0$ is the critical chemical potential $\mu_c=\mu_0$.

At the zeroth order, $\phi_0$ can be obtained from eq.(\ref{eom22}) as

\begin{equation} \phi_0(u)=\lambda r_{+c}^{-2}\log{u} \label{phi0}\end{equation}

where we have introduced a new parameter $\lambda=\mu_c r_{+c}^2$. On the other hand, from the asymptotical behavior of the scalar field, we can define a trial function $F(u)$

\begin{equation} \psi(u)=\epsilon r_+^{\Delta} u^{\Delta} F(u) \label{trial}\end{equation}

with proper boundary conditions for $F(u)$ to match the behavior of the scalar field

\begin{equation} F(0)=1\ ,\qquad F'(0)=0 \label{bdyf}\end{equation}

From eq.(\ref{eom11}) and eq.(\ref{trial}), we deduce

\begin{equation} F''(u)+\frac{T'(u)}{T(u)}F'(u)+[\lambda^2 V(u)+U(u)]F(u)=0 \label{eomtrial}\end{equation}

with

\begin{equation} T(u)=u^{\Delta+1} h(u)  \end{equation}

\begin{equation} V(u)= \frac{u^2\log^2{u}}{h^2(u)} \end{equation}

\begin{equation}  U(u)=-\frac{\Delta^2 u^{\Delta-2}+\tilde{m}^2 u^{2-\Delta}}{h(u)}\end{equation}

Using Sturm-Liouville eigenvalue method, we are readily to derive an expression to estimate the minimum value of $\lambda^2$

\begin{equation} \lambda^2=\frac{\int_0^{1}\mathrm{d}u [F'^2(u)-U(u)F^2(u)]}{\int_0^{1}\mathrm{d}u T(u)V(u)F^2(u)} \label{lambda} \end{equation}

To explicitly applying the variational method, we assume the trial function to be $F(u)=1-a u^2$, where $a$ is an undetermined coefficient. As an example, for $\theta=1$, we obtain

\begin{equation} \lambda^2=\frac{1.5-2.25 a+1.071428 a^2}{0.01031-0.011416 a+0.0036062 a^2}\ ,\quad \mathrm{for}\ \  m^2=0  \end{equation}

\begin{equation} \lambda^2=\frac{0.75-1.25 a+0.69643 a^2}{0.01031-0.011416 a+0.0036062 a^2}\ ,\quad \mathrm{for}\ \  \tilde{m}^2=-3  \end{equation}

The critical temperature measured in units of the chemical potential is related to the minimum of $\lambda$ as

\begin{equation} T_c=\frac{\Delta}{4\pi \lambda_{min}} \mu \label{tc}\end{equation}

\FIGURE[ht]{\label{figure3}
\includegraphics[width=7.5cm]{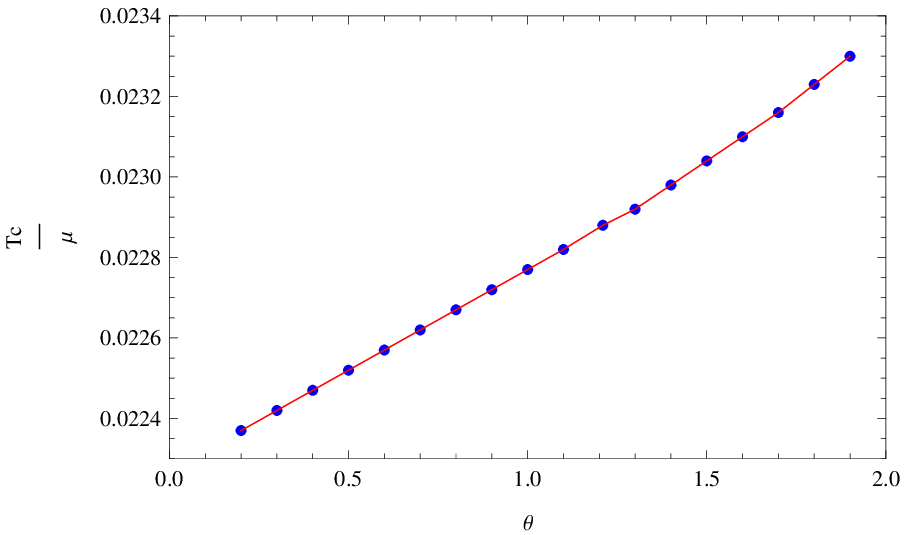}
\includegraphics[width=7.5cm]{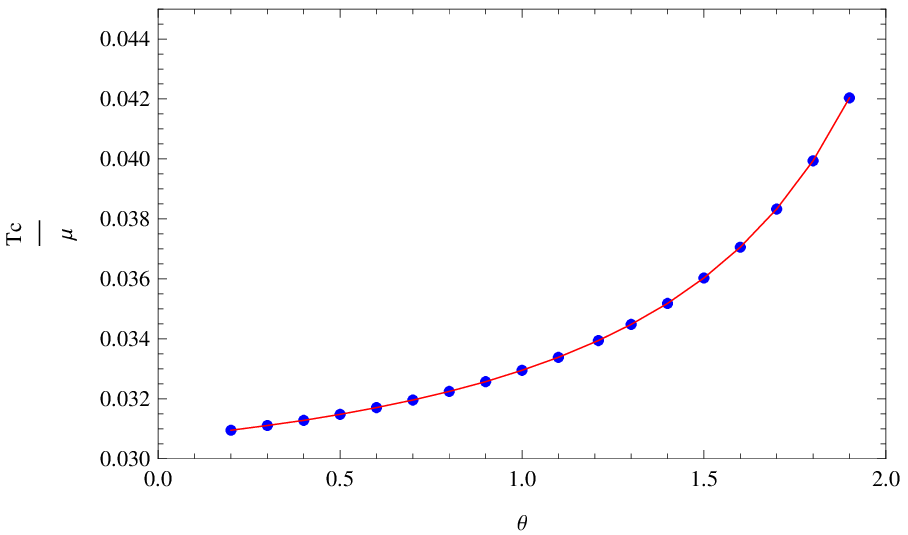}
\caption{The dependence of the critical temperature $T_c/\mu$ on the hyperscaling violation. The left plot for $m^2=0$, the right plot for $\tilde{m}^2=-3$.}
}

We find that $T_c=0.0229604 \mu$ for $m^2=0$ and $T_c=0.032948 \mu$ for $\tilde{m}^2=-3$, which are in excellent agreement with our previous numerical results. In figure \ref{figure3}, we plot the critical temperature with various value of hyperscaling violation $\theta \in (0,2)$, allowed by eq.(\ref{bf}). From the left plot, we see that the ratio $T_c/\mu$ increases roughly linearly\footnote{This is not true in  a strict sense. Actually, from the $\theta=1.3$ point, there exists a little slope increasing, but very slowly with the increasing of $\theta$. } as $\theta$ increases while in the right plot it also increases but not linearly again. This is not surprising. Since the fixed parameter is $m^2$ for the left plot while in the right plot it is $\tilde{m}^2=m^2 r_{+c}^\theta$, the true mass square at the different points changes with the critical temperature variation, leading to the different behavior from the left plot. The increase of the critical temperature with the increasing of the hyperscaling violation indicates that the scalar condensate becomes easier for bigger value of the hyperscaling violation. Thus, the nontrivial role of the hyperscaling violation in the superconducting phase transition is probably to provide some unknown mechanism to produce more net attraction force between electrons such that Cooper pairs emerges a little easier in the systems with higher hyperscaling violation. In order to get a better understanding in a deeper level at this point, we need to extract the behavior of the low energy degrees of freedom near the superconducting transition and provide an effective way to increase the critical temperature for general superconductors. This is clearly the central topic in the studies of high temperature superconductors. However, a perfect claim and an effective theory framework are still lacking. From the analysis of entropy balance\cite{hartnoll} in holographic superconductors, recently given by S.A. Hartnoll and R. Pourhasan, a naive physical argument is that the increasing of the critical temperature probably depends on the increasing of the low energy degree of freedom in the normal state relative to the low energy degree of freedom in the superconducting state. It is certainly interesting to extract the detail of how the hyperscaling violation works in this sense. We hope to return to this subject in the near future.

In order to ensure the above calculation for the critical temperature is physically reasonable, we also need to demonstrate the scalar condensate behaves well near the critical point, without any oscillation and instability. This is equivalent to claim the chemical potential correction $\delta \mu_2$ is always positive definite for arbitrary hyperscaling violation in the region $(0,2)$. We find this is true and present the detail in Appendix.

\section{Conclusions}

In this paper, we have investigated holographic superconductors in asymptotically geometries with hyperscaling violation, a natural generalization of the Lifshitz geometry with zero entropy density at the zero temperature limit. This is the most general effective holographic model of strongly coupled condensed-matter theory. We focus on  $z=2$ Lifshitz scaling and first set the hyperscaling violation exponent $\theta=1$ to numerically solve the non-linear coupled Maxwell-dilaton equations of motion in the probe limit. We find that the scalar condenses when the temperature crosses a critical value but does not approaches a fixed constant at the low temperature limit, indicating that the backreaction effects should be included when the temperature is sufficiently lowered. Furthermore, the real part of the conductivity approaches 1 in the high frequency limit where the condensate can be neglected in the fluctuations equation, consistent with the normal phase. And a gap opens near the zero frequency and gets deeper as the temperature further lowered. In the imaginary part of the conductivity, there exists a pole, implying that a delta function is contained in the real the part of the conductivity according to the Kramers-Kr\"{o}nig relation.

By applying the Sturm-Liouville eigenvalue method, we further analytically explore the effects of the hyperscaling violation on the superconducing phase transition. The critical temperature was shown to increase almost linearly when $\theta$ increases for fixed mass square of the scalar field. This enhancement effect probably indicates that the system with hyperscaling violation produces more net effective attraction between electrons. The underlying mechanism remains mysterious at the present stage.

\section{Appendix}

Substitute the expansion of the gauge field eq.(\ref{expansion}) into the equations of motion eq.(\ref{eom22}), we obtain

\begin{equation} \phi_2''(u)+\frac 1u \phi_2'(u)=-\lambda \frac{H(u)}{u} \label{newphi}\end{equation}

where we have scaled the field as $\phi_2(u)\rightarrow r_+^{\Delta+2}\phi_2(u)$ such that the gauge field is expanded as $\phi(u)=\phi_0(u)+\epsilon^2r_+^{\Delta+2}\phi_2(u)$. The function $H(u)$ is given by

\begin{equation}H(u)=\frac{-2u^{\Delta+3}\log{u}F^2(u)}{h(u)} \end{equation}

Solve eq.(\ref{newphi}) with boundary conditions $\phi_2(1)=\phi_2'(1)=0$, we find

\begin{equation} \phi_2(u)\mid_{u\rightarrow 0}\approx\lambda \mathcal{C}\log{u}\ ,\qquad \mathcal{C}=\int_0^1 \mathrm{d}u H(u)  \end{equation}

Recall that the gauge field $\phi(u)$ asymptotically behaves as $\phi(u)\mid_{u\rightarrow 0}\sim \mu \log{u}$, we finally obtain

\begin{equation} \mu\approx \mu_c+\epsilon^2 \delta\mu_2\ ,\qquad \delta\mu_2=\mu_cr^{2\Delta+\theta}_{+c}\mathcal{C} \end{equation}

Since the function $H(u)$ is always positive in the region $u\in(0,1)$, its integral i.e. the constant $\mathcal{C}$ is positive. Thus we confirm that the chemical potential correction $\delta\mu_2$ is always positive definite, independent of the hyperscaing violation.
\section{Acknowledgments}
I thank Dr.Yanyan Bu very much for his help in mathematical codes and useful discussions. I also thank Professor Sije Gao for his great encouragement. This work is supported by NSFC Grants NO.10975016, NO.11235003 and NCET-12-0054.

\end{document}